# *Dynamical Theory of Artificial Optical Magnetism Produced by Rings of Plasmonic Nanoparticles*

Andrea Alù, and Nader Engheta[*]

University of Pennsylvania, Dept. of Electrical and Systems Engineering, Philadelphia,

19104 PA, U.S.A.

We present a detailed analytical theory for the plasmonic nanoring configuration first proposed in [A. Alù, A. Salandrino, N. Engheta, Opt. Expr. **14**, 1557 (2006)], which is shown to provide negative magnetic permeability and negative index of refraction at infrared and optical frequencies. We show analytically how the nanoring configuration may provide superior performance when compared to some other solutions for optical negative index materials, offering a more "pure" magnetic response at these high frequencies, which is necessary for lowering the effects of radiation losses and absorption. Sensitivity to losses and the bandwidth of operation of this magnetic inclusion are also investigated in details and compared with other available setups.

1. **Introduction**

Engineering metamaterials with anomalous values of their constitutive parameters, and in particular with negative real parts of their permittivity $\varepsilon$ and/or permeability $\mu$, has become a popular topic of research interest since the anomalous refractive properties of left-handed materials (i.e., media with both effective permittivity and permeability with negative real part and sufficiently low losses) have been demonstrated experimentally at microwave frequencies [1]. The technology behind such composite materials is fairly well established in this frequency regime and involves the use of resonant man-made inclusions, with sizes much smaller than the wavelength of operation, interacting with the local electromagnetic fields to induce a proper resonant response in the effective permittivity and/or permeability of the composite homogenized metamaterial. At these relatively low frequencies this is commonly obtained by utilizing conducting materials shaped as dipoles, that induce an $\varepsilon$-negative response [2], and as split-ring resonators (SRR), that provide negative real part for $\mu$ [3].

---

[*] To whom correspondence should be addressed. E-mail: engheta@ee.upenn.edu

The interest in translating these concepts to higher frequencies, namely near infrared and optical frequencies, has met some technological challenges for what concerns the engineering of a negative real part of permeability. If some plasmonic materials, like noble metals and polar dielectrics, have naturally a negative permittivity around this range of frequencies [4], no natural materials indeed possess a negative permeability at optical frequencies. The same concept of permeability has even been questioned at optical frequencies in highly respected textbooks [5], §79, where it is argued that when the atomic size of the materials becomes comparable with the wavelength of the electrons, i.e., for sufficiently high frequencies, a magnetic polarization cannot be associated with the electromagnetic response of materials and it is "an over-refinement to distinguish between the magnetic displacement **B** and the magnetic field **H**". It is worth noting that in the same book it is also stated that at optical frequencies "the effects due to the difference of µ from unity are in general indistinguishable from those of the spatial dispersion of the permittivity" [5], §103, implicitly assuming that although a magnetic response in its classic sense is not expected from optical frequencies onwards, a proper spatial dispersion in the material response may effectively produce a magnetic response noticeably different from that of free space. This is clearly related to the well known concept, true at any frequency range, that Maxwell equations may be equivalently written by embedding the magnetic polarization into a proper spatial dispersion of the permittivity [5]-[6]. In other words, it appears possible, at least in principle, to envision a proper sub-wavelength molecular shape interacting with the optical magnetic field (or, if preferred, with the curl of the optical electric field) inducing a non-negligible magnetic dipole moment.

If a mere scaling of the SRR inclusion, widely employed at lower frequencies, is not easily feasible due to the change in the conduction properties of optical materials and the related well known problems of saturation [7], several proposals have been presented in the recent years to overcome this difficulty and engineering an effective µ-negative metamaterial in the visible. In [8] an SRR-shaped metallic inclusion has been shown to work for this purpose up to the THz (infrared) regime, whereas in [9]-[10] properly engineered defects in photonic crystals have been shown to induce a dominant magnetic dipole moment at optical frequencies. Guided waves in anisotropic or plasmonic waveguides have been shown to "experience" an effective negative permeability [11]-[13], and many other proposals of utilizing properly designed plasmonic resonances for inducing a magnetic dipole response have been presented independently by several groups [14]-[26].

In particular, the coupling between a pair of closely spaced plasmonic nanoparticles has been shown in [14]-[25] to support an anti-symmetric resonance exhibiting a sensible magnetic dipole contribution. In order to make this contribution significant, however, in some of these works the size of the inclusion had to be comparable with the wavelength of operation, making hardly distinguishable the metamaterial resonance due to the particles embedded in the material from the lattice resonance typical of any photonic band-gap structure. Even the homogenization procedure for assigning an effective permeability may be inadequate when the lattice period becomes comparable with the wavelength of operation [27]-[29].

Coming from a different point of view, for the same purposes we have proposed to exploit the resonance supported by arrangements of (more than two) plasmonic nanoparticles placed around a circular sub-wavelength loop [26], in order to synthesize a nanoring magnetic resonator at optical frequencies. We have shown how this geometry effectively responds to a magnetic field excitation with a dominant magnetic dipole response, similar to what an SRR inclusion would do at lower frequencies. The main difference consists in the physics behind the two phenomena: if an SRR at microwave frequencies supports a resonant *conduction* current circulating around the center of the inclusion following the specific shape of the loop and guided by the conducting material composing the ring, in the nano-ring geometry the *displacement* current takes the same role, being guided by the plasmonic resonances of the particles composing the loop. If the conduction of materials is weakened by an increase of the frequency of operation ($e^{-i\omega t}$ dependence), the role of the displacement current, i.e., the time derivative of the electric displacement vector $-i\omega \mathbf{D}$, being proportional to the frequency of operation, indeed takes a dominant role at these high frequencies. The proper arrangement of such plasmonic nano-particles, each of them near their own resonant frequency, around a sub-wavelength loop supports a resonant circulation of displacement current that induces a strong magnetic response of the nanoring in its entirety. This interpretation is also consistent with our recently introduced nanocircuit paradigm [30]-[31], for which loops of plasmonic nanoparticles interleaved by insulating gaps may be interpreted as the resonant interconnection of inductors and capacitors in a loop shape, somewhat transplanting the concept of SRR into the optical nanocircuits.

An important point arising from our analysis [26] consists in the fact that the size of the nanoring is weakly related to the resonant properties of this magnetic effect, but rather its resonance is mainly associated with the plasmonic features of the individual particles constituting the nano-ring. Its size, therefore, may be designed to be sufficiently small to allow the proper definition of an effective permeability for an optical metamaterial constituted of such nano-ring inclusions.

In [26] it was also shown how this same inclusion may possess ,in addition, a resonant electric response and, provided that the electric and magnetic resonances are designed to arise at close frequencies, the resulting metamaterial may have overall left-handed properties, combining a negative real part of its effective permittivity and permeability. This may constitute an interesting venue to design left-handed metamaterials at optical frequencies.

In this paper, following these concepts, we develop a fully dynamic analysis of the magnetic response of the nanoring geometry, providing physical insights into this anomalous magnetic nanoresonance and comparing its magnetic response with the one obtained from pairs of plasmonic nanoparticles. We will show how the use of multiple particles arranged in this circular geometry ensures a stronger and "cleaner" magnetic dipole response, effectively creating the first attempt to design a purely magnetic inclusion at optical frequencies. We also study in detail the bandwidth of such resonance and the sensitivity to losses as a function of the geometry of this novel setup.

## 2. Geometry of the problem

As proposed in [26], our geometry of interest is formed by a collection of $N$ plasmonic sub-wavelength isotropic nanoparticles of polarizability $\alpha_p$ arranged equidistantly around a circumference of radius $R \ll \lambda_b$ in the $x$-$y$ plane, as schematically depicted in Fig. 1. Here $\lambda_b$ is the wavelength of operation in the background material, which is assumed in the following to have a permittivity $\varepsilon_b$ and free-space permeability $\mu_0$. The excitation is assumed monochromatic with $e^{-i\omega t}$ time dependence. If the center of the loop is at the origin of a Cartesian reference system, the center of each particle is placed symmetrically with respect to the origin at the point $\mathbf{r}_j \equiv \left( R\cos(2\pi j/N), R\sin(2\pi j/N), 0 \right)$ with $j = 0...N-1$. Notice how this general geometry may describe also the situation of a pair of two closely spaced nanoparticles, as in [14]-[22], which would correspond to the case of $N = 2$, with their center-to-center distance being equal to $2R$.

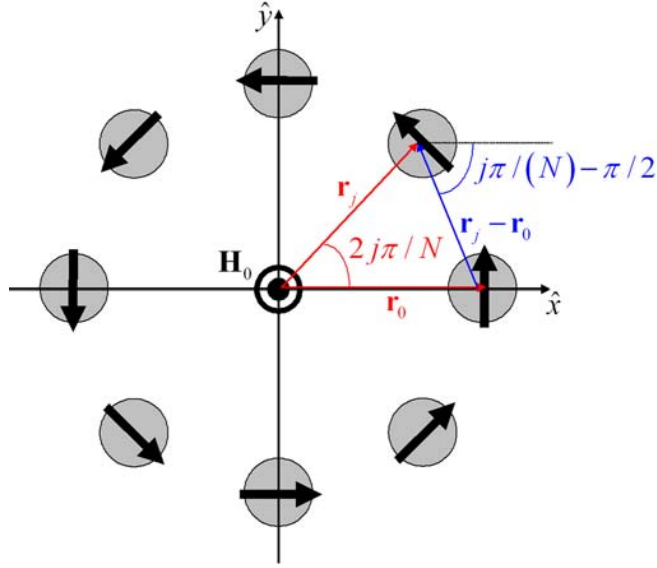

Fig. 1 – A nano-ring made of $N$ nanoparticles symmetrically displaced around the origin of a Cartesian reference system in the $x$-$y$ plane. The electric polarization response to a uniform magnetic excitation is depicted in the figure, ensuring purely rotational induced electric dipoles (black arrows).

### 3. Magnetic response of the nanoring

Consider now the response of this nanoring to a uniform magnetic field excitation $\mathbf{H}_{imp} = H_{imp}\hat{\mathbf{z}}$ directed along the axis of the loop. The hypothesis of describing the electromagnetic interaction of the particles composing the loop only through their electric polarizability $\alpha_p$ is justified by the sub-wavelength size of each one of the particle and by the fact that they are non-magnetic, i.e., the permeability of each particle and of the background material are all equal to that of free space $\mu_0$. This implies that each of the particles responds solely to the local electric field $\mathbf{E}_{loc}$ and not directly to $\mathbf{H}_{imp}$. Integrating Maxwell equation $\nabla \times \mathbf{E}_{imp} = i\omega\mu_0\mathbf{H}_{imp}$ under the assumptions of a uniform quasi-static magnetic field over the volume occupied by the loop and of sub-wavelength dimensions, we find that the following relation holds on the circumference where the particles are placed:

$$\mathbf{E}_{imp} = \frac{i\omega\mu_0 R H_{imp}}{2}\hat{\boldsymbol{\varphi}}. \qquad (1)$$

Eq. (1) derives the amplitude of the electric field impressed on each particle, relating the magnetic response of the nanoring to the local variation, i.e., the curl, of the impressed electric field, consistent with the previous discussion on the necessity of tailoring the weak spatial dispersion of the molecule to achieve, at these high-frequencies, a non-negligible magnetic response. It should be noted that under the quasi-static assumption of a uniform magnetic field on the nano-loop, the averaged electric field on the system is zero, as it is clear from (1). This allows isolating the magnetic response of the nanoring from its possible electric response, which has been analyzed numerically in [26].

Each of the nanoparticles composing the loop, therefore, is excited in this case by an impressed electric field directed along the tangent to the circumference along which they are positioned, and proportional to the uniform magnetic field on the loop. Owing to the symmetry of geometry and excitation, the electric dipoles induced over the particles are also directed along $\hat{\boldsymbol{\varphi}}$, as depicted in Fig. 1 by the black arrows over each particle. The induced dipole amplitude is proportional to the local electric field induced on the volume of each particle when its self-polarization contribution is not considered [27]-[28], which is $\mathbf{E}_{loc} = \mathbf{E}_{imp} + \sum_{j \neq j'} \mathbf{E}_{jj'} = E_{loc}\hat{\boldsymbol{\varphi}}$, through the proportionality factor $\alpha_p$. In this expression $\mathbf{E}_{jj'}$ is the electric field induced by the dipole $j$ on the position where the particle $j'$ is placed. Each one of the $N$ particles can, therefore, be represented as an effective dipole moment $\mathbf{p} = p\hat{\boldsymbol{\varphi}} = \alpha_p E_{loc}\hat{\boldsymbol{\varphi}}$. Since $\mathbf{E}_{jj'} = p\underline{\mathbf{Q}}_{jj'} \cdot \hat{\boldsymbol{\varphi}}(\mathbf{r}_{j'})$, where:

$$\mathbf{\underline{Q}}_{jj'} = \frac{e^{ik_b|\mathbf{r}_j - \mathbf{r}_{j'}|}}{4\pi\varepsilon_b |\mathbf{r}_j - \mathbf{r}_{j'}|} \left\{ k_b^2 \left[ \mathbf{\bar{I}} - \mathbf{\bar{D}}_{jj'} \right] + \left( \frac{1}{|\mathbf{r}_j - \mathbf{r}_{j'}|^2} - \frac{ik_b}{|\mathbf{r}_j - \mathbf{r}_{j'}|} \right) \left[ 3\mathbf{\bar{D}}_{jj'} - \mathbf{\bar{I}} \right] \right\} \quad (2)$$

is the 3-D dyadic Green's function as usually defined [27], with $\mathbf{\bar{D}}_{jj'} = \frac{\mathbf{r}_j - \mathbf{r}_{j'}}{|\mathbf{r}_j - \mathbf{r}_{j'}|} \frac{\mathbf{r}_j - \mathbf{r}_{j'}}{|\mathbf{r}_j - \mathbf{r}_{j'}|}$,

$\mathbf{\bar{I}}$ being the identity dyadic, $k_b = \omega\sqrt{\varepsilon_b \mu_0} = 2\pi/\lambda_b$ and $\hat{\boldsymbol{\varphi}}(\mathbf{r}_{j'})$ is the spherical unit vector $\hat{\boldsymbol{\varphi}}$ at the location $\mathbf{r}_{j'}$, the final closed-form expression for $p$ is given by:

$$p = \frac{i\omega\mu_0 R H_{imp}/2}{\alpha_p^{-1} - \sum_{j \neq j'}^{N-1} \mathbf{\underline{Q}}_{jj'} \cdot \hat{\boldsymbol{\varphi}}(\mathbf{r}_j) \cdot \hat{\boldsymbol{\varphi}}(\mathbf{r}_{j'})}. \quad (3)$$

This expression, although derived under a different form of excitation, is consistent with the results in [26], and it is a fully dynamic expression valid within the only approximation that the dominant multipole order for each nanoparticle composing the loop is dominated by its electric dipole response.

For expressing the summation in the denominator of (3) in a compact form, in [26] we have used the static approximation of $\mathbf{\underline{Q}}_{jj'}$, which neglected the radiation contribution from the individual dipoles and therefore did not satisfy the energy conservation relations for the nanoring. Here, instead, we analyze in more details the dynamic expression (3), which allows us to show and discuss what are the inherent fundamental limits and the conditions under which such nanoring may be regarded as an effective magnetic inclusion.

For symmetry, the summation in (3) is independent of the index $j'$, and therefore in its evaluation we can consider $j' = 0$ without losing generality. This implies that we are evaluating the specific effective dipole amplitude of the particle placed at the point $\mathbf{r}_0 = (R, 0)$. In this dynamic case, the expression inside the summation in (3) may be evaluated in closed form by noticing that $|\mathbf{r}_j - \mathbf{r}_0| = 2R \left| \sin \dfrac{j\pi}{N} \right|$ and the angle $\beta$ formed by $\mathbf{r}_j - \mathbf{r}_0$ and $\hat{\mathbf{x}}$ is $\beta = \dfrac{j\pi}{N} - \dfrac{\pi}{2}$, as geometrically depicted in Fig. 1. We therefore obtain the interesting dynamic result:

$$\underline{\mathbf{Q}}_{j0} \cdot \hat{\boldsymbol{\varphi}}(\mathbf{r}_j) \cdot \hat{\boldsymbol{\varphi}}(\mathbf{r}_0) =$$
$$= \frac{e^{2ik_b R \sin(j\delta)}}{64\pi R^3 \varepsilon_b} \frac{3 - 3k_b^2 R^2 + (1 + 4k_b^2 R^2)\cos(2j\delta) - k_b R \{k_b R \cos(4j\delta) + i[5\sin(j\delta) + \sin(3j\delta)]\}}{\sin^3(j\delta)}$$

, (4)

with $\delta = \pi / N$.

Expanding this expression in the Taylor series of $R$, we can have a better insight into the meaning of Eq. (3). In particular, the imaginary part of (4) contains only even powers of $R$, whereas the odd powers are associated with its real part. We can write:

$$\operatorname{Re}\left[\underline{\mathbf{Q}}_{j0} \cdot \hat{\boldsymbol{\varphi}}(\mathbf{r}_j) \cdot \hat{\boldsymbol{\varphi}}(\mathbf{r}_0)\right] = \frac{3 + \cos(2j\delta)}{64\pi \varepsilon_b R^3 \sin^3(j\delta)} + o(k_b R)$$
$$\operatorname{Im}\left[\underline{\mathbf{Q}}_{j0} \cdot \hat{\boldsymbol{\varphi}}(\mathbf{r}_j) \cdot \hat{\boldsymbol{\varphi}}(\mathbf{r}_0)\right] = \frac{k_b^3 \cos(2j\delta)}{6\pi \varepsilon_b} + \frac{k_b^5 R^2 [1 - 3\cos(2j\delta)] \sin^2(j\delta)}{30\pi \varepsilon_b} + o(k_b R)^4$$

. (5)

The imaginary part in (5), in particular, relates directly to the power radiated by each one of the dipoles, providing an insight into the dynamic physical behavior of the nanoring (this imaginary part had been neglected in our quasi-static analysis in [26]).

It can be noted from Eq. (3) that a self-sustained (i.e., source-free, for $H_{imp} = 0$) eigensolution for such a system of particles may be obtained under the condition that the denominator in (3) vanishes, i.e.:

$$\sum_{j=1}^{N-1} \underline{\mathbf{Q}}_{j0} \cdot \hat{\boldsymbol{\varphi}}(\mathbf{r}_j) \cdot \hat{\boldsymbol{\varphi}}(\mathbf{r}_0) = \alpha_p^{-1}. \tag{6}$$

The complex dispersion relation (6) for the nanoring resonances depends directly on the inverse of the polarizability $\alpha_p^{-1}$ of each nanoparticle in the loop. This quantity represents just one degree of freedom in our problem, represented by its real part (which corresponds to the reactance of our particle), since $\text{Im}\left[\alpha_p^{-1}\right]$ is directly related to the radiation and ohmic losses of the particle itself and it is determined merely by energy relations, consistent with the discussion for the linear array problem that we have discussed in [32]. For instance, it is well known how for a single isolated particle with lossless material, $\text{Im}\left[\alpha_p^{-1}\right] = -k_b^3/(6\pi\varepsilon_b)$, due to its dipolar scattering loss [28].

This implies that a self-sustained resonant condition may not be achievable for passive nanoparticles, since, even in the ideal lossless case, they radiate some energy. A complete self-sustained resonance may be obtained only if the particles may compensate such radiation, i.e., if they are "active". This is confirmed, taking the imaginary part of both sides of (6), using (5) and evaluating the involved summation. For any $N > 2$ this yields:

$$-\frac{k_b^3}{6\pi\varepsilon_b} + \frac{N k_b^5 R^2}{24\pi\varepsilon_b} + o\left[(k_b R)^4\right] = \text{Im}\left(\alpha_p^{-1}\right). \tag{7}$$

For passive particles, for which [32]:

$$\text{Im}\left[\alpha_p^{-1}\right] = -k_b^3/(6\pi\varepsilon_b) - \alpha_{loss}, \tag{8}$$

with $\alpha_{loss}$ being a strictly positive quantity taking into account the possible ohmic absorption, Eq. (7) cannot be satisfied, and a self-sustained mode is not achievable due to the inevitable presence of radiation losses. In particular, it is interesting to note how, in the limit of lossless particles, or whenever ohmic absorption may be neglected, the right hand side of (7) cancels out the first term in the left hand side for any $N \geq 2$, term that corresponds to the contribution to the electric dipole moment radiation. Physically, this is explained by the fact that a system with more than one particle arranged in this configuration and excited by a uniform magnetic field does not possess any electric dipole moment. Moreover, as we show in the following, the second term in (7), which is a strictly positive quantity, takes into account the power associated with the magnetic dipole radiation, whereas the $o\left[(k_b R)^4\right]$ is associated with the higher-order multipole radiation contributions. Considering the presence of these radiative terms (i.e., considering the dynamic behavior of the Green's function) implies the finiteness of the amplitude in (3) for any (real) frequency, as a symptom of the presence of radiation damping from the loop, even in the absence of Ohmic absorption.

After these considerations, it is interesting to expand the current distribution represented by these induced dipoles, all directed along $\hat{\boldsymbol{\varphi}}(\mathbf{r}_j)$ and with equal amplitude (3), in terms of its multipole moments. In other words we evaluate here the quasi-static multipole expansion of the current density $\mathbf{J} = -i\omega p \sum_{j=0}^{N-1} \hat{\boldsymbol{\varphi}}(\mathbf{r}_j) \delta(\mathbf{r}-\mathbf{r}_j)$. The quasi-static electric multipoles of order $n$, indicated in the following as $\underline{\mathbf{p}}_H^{(n)}$, with the subscript $H$ to indicate that they are generated by a uniform magnetic excitation of the loop, may be written as the following $n-$adic [33], §4.1:

$$\underline{\mathbf{p}}_H^{(n)} = \frac{i}{\omega}\left(\int_V \mathbf{J}\mathbf{r}^{(n-1)} + \mathbf{r}\mathbf{J}\mathbf{r}^{(n-2)} + \ldots + \mathbf{r}^{(n-2)}\mathbf{J}\mathbf{r} + \mathbf{r}^{(n-1)}\mathbf{J}\right)dV =$$
$$= p\sum_{j=0}^{N-1} \hat{\boldsymbol{\varphi}}(\mathbf{r}_j)\mathbf{r}_j^{(n-1)} + \ldots + \mathbf{r}_j^{(n-1)}\hat{\boldsymbol{\varphi}}(\mathbf{r}_j) \quad , \tag{9}$$

where $V$ is any volume containing the nanoparticles composing the loop, and $\mathbf{r}^{(n-1)}$ is the $(n-1)$-adic formed by the consecutive dyadic products of $(n-1)$ vectors $\mathbf{r} = (x, y, z)$.

Expanding the summation in (9) and applying geometrical considerations, it follows that:

$$\underline{\mathbf{p}}_H^{(n)} = \underline{\mathbf{0}} \qquad \text{for any } n \leq N-1, \tag{10}$$

and the amplitude of the residual non-vanishing higher-order electric multipoles is proportional to $R^{n-1}$ [34].

This result implies, as anticipated, that any configuration with two or more identical nano-particles displaced symmetrically around the origin does not show any electric dipole moment response to a uniform magnetic excitation. The higher is the number of particles, the more electric multipoles are canceled.

The magnetic multipoles $\underline{\mathbf{m}}_H^{(n)}$ may be written by duality using formula (9) after the substitution $\mathbf{J} \to -\dfrac{i\omega}{2} \mathbf{r} \times \mathbf{J}$ [33]. In this case, applying similar geometrical considerations it follows that:

$$\underline{\mathbf{m}}_H^{(2n)} = \underline{\mathbf{0}} \qquad \text{for any } 2n \leq N, \tag{11}$$

whereas odd magnetic multipoles $\underline{\mathbf{m}}_H^{(2n+1)}$ are always present in the radiation from the nano-ring. In particular, the magnetic dipole moment has amplitude:

$$\mathbf{m}_H^{(1)} = \dfrac{-i\omega p N R}{2} \hat{\mathbf{z}}, \tag{12}$$

and the amplitude of all the non-vanishing magnetic multipoles is proportional to $N R^n$.

The result in (12) is consistent with the findings derived heuristically in [26] by analyzing the far-field radiation by nanoring in the limit of $(k_b R) \to 0$. This multipole expansion rigorously proves this result and aims to address the extent to which its electromagnetic response may be effectively considered "purely" magnetic as a function of its geometrical parameters. In other words, in the following we answer the questions: 'How well may this configuration of nanoparticles support a pure magnetic resonance as a function of its geometry and number of particles?', and 'How close may the near-field resemble the one produced by an ideal magnetic loop?'. The answer to these questions is important when we aim to embed such nanorings as artificial molecules in a host medium in order to form a bulk metamaterial with magnetic and/or negative index properties. We show in the following the advantages of employing a larger number of particles around the loop. When the particles composing the nanoring are only two, the radiated field is dominated not only by the magnetic dipole moment, with expression given by (12) $\mathbf{m}_H^{(1)} = -i\omega pR\hat{\mathbf{z}}$, but also by a non-negligible electric quadrupole moment $\mathbf{p}_H^{(2)} = 2pR(\hat{\mathbf{x}}\hat{\mathbf{y}} + \hat{\mathbf{y}}\hat{\mathbf{x}})$. This quadrupolar contribution, which is often neglected in the analysis of metamaterials constituted by pairs of coupled nanoparticles, has been noticed in [35]-[36]. The magnetic and quadrupolar contributions, as evident from the previous formulas, are of the same order with respect to $R$, and therefore the quadrupolar contribution cannot be reduced by varying the nanopair size. This should be clearly taken into account when such pairs are embedded in a bulk metamaterial. At the other extreme, when the number of nano-particles becomes large, the electric multipoles are all negligible, whereas only the odd magnetic multipoles have a role in the radiation from such loop. For a sufficiently small loop, however, the contribution of the magnetic dipole moment, equal to (12), dominates the second major contribution, i.e., the magnetic octupole moment, which is equal to $\mathbf{m}_H^{(3)} = -i\omega pNR^3/2(\hat{\mathbf{x}}\hat{\mathbf{x}}\hat{\mathbf{z}} + \hat{\mathbf{x}}\hat{\mathbf{z}}\hat{\mathbf{x}} + \hat{\mathbf{z}}\hat{\mathbf{x}}\hat{\mathbf{x}} + \hat{\mathbf{y}}\hat{\mathbf{y}}\hat{\mathbf{z}} + \hat{\mathbf{y}}\hat{\mathbf{z}}\hat{\mathbf{y}} + \hat{\mathbf{z}}\hat{\mathbf{y}}\hat{\mathbf{y}})$. When the number of particles is not large enough and/or the loop is not electrically very small, so that higher-order multipoles may be non-negligible, it is preferable to use an even number of particles, in order to cancel the contribution from even magnetic and odd electric multipoles (see [34]-[37]).

Following this discussion, Fig. 2 shows the calculated field distribution for the electric field $E_\phi$ in the plane $\theta = \pi/2$ at different radial distances $r$ from the origin, normalized to the (uniform) field that would be radiated by an ideal magnetic dipole with the amplitude given by (12). The different examples are calculated by varying the number of particles and the size of the loop. We note that when the nanoring is electrically very small and there is a sufficiently large number of particles composing the loop ($R = \lambda_b/100$, $N = 6$, Fig. 2a), it is sufficient to be at a distance of 10 times the particle's radius (still very close to the loop, at a distance $r = \lambda_b/10$) to experience a field distribution identical to that of a pure magnetic dipole, and the amplitude of the normalized field rapidly converges to unity. As we increase the dimension of the loop for the same number of particles ($R = \lambda_b/10$, $N = 6$, Fig. 2b), the deviation from the magnetic dipole field distribution becomes less sensitive to the relative distance from the origin (normalized to the radius size), since we are physically more distant from each nano-particle. This is consistent with the fact that, although the amplitude of higher-order multipoles increases for larger $(k_b R)$, their radiated fields decay faster away from the origin. Reducing the number of spheres ($N = 4$, Fig. 2c-d), the distance plays a more significant role, since lower order multipoles start to contribute significantly (notice also the difference of scales in the figures 2c and 2e). Reducing the loop to just a pair of particles, as in Fig. 2e-f, the field distribution in the near as well as the far-field is drastically distorted by the presence of the electric quadrupole radiation, which affects dramatically the shape of the scattered field from the pair, independently of its size. It is also worth noting that in the three cases with bigger loops (Figs. 2b,d,f) the normalized far-zone field does not converge exactly to unity, as it happens for the smaller loop case of Fig. 2a or 2c, since higher-order multipoles also start to contribute noticeably to the far-zone field.

This analysis clarifies the extent to which resonant nanopairs may be effectively considered and interpreted as resonant "magnetic dipoles". It is clear how, in the setups proposed in [14]-[22] employing nanopairs to design optical negative-index metamaterials, special attention should be paid in interpreting the resonant response of the pair as a magnetic dipole, since the quadrupolar contribution from such pairs cannot be neglected and it is instead comparable with the magnetic response. As we discuss in the following, these serious effects should be properly taken into account in the proper homogenization of a metamaterial formed by such inclusions [35].

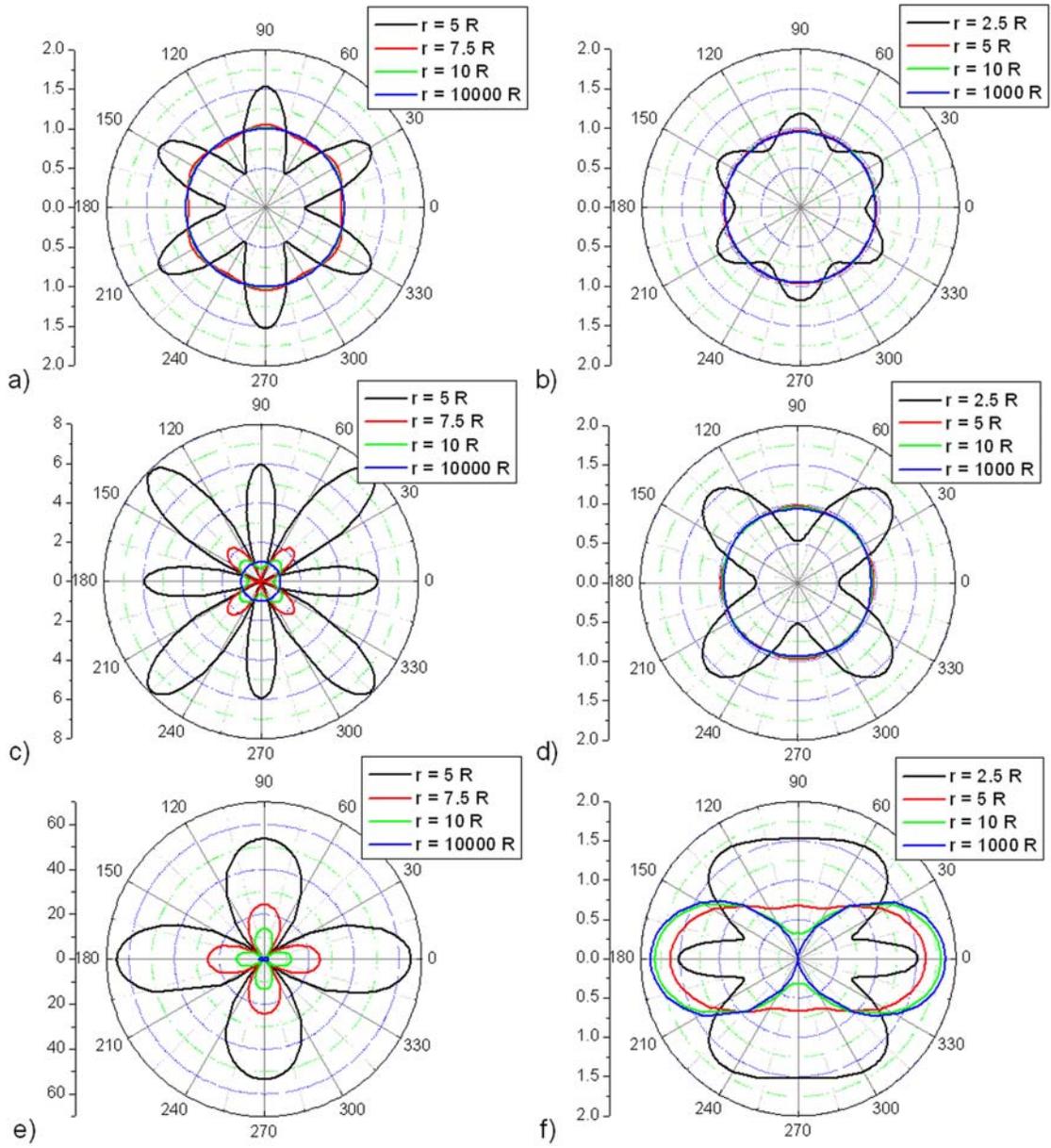

Fig. 2 – Electric field distribution in the $\theta = \pi/2$ ($x$-$y$) plane varying the distance $r$ from the origin, for the resonant loop of Fig. 1 with: a) $R = \lambda_b/100$, $N = 6$; b) $R = \lambda_b/10$, $N = 6$; c) $R = \lambda_b/100$, $N = 4$; d) $R = \lambda_b/10$, $N = 4$; e) $R = \lambda_b/100$, $N = 2$; f) $R = \lambda_b/10$, $N = 2$.

Having evaluated the magnetic dipole (12) induced by a uniform magnetic field impressed on the configuration of Fig. 1, it is possible to define the magnetic polarizability $\alpha_m$ of this nano-loop, which satisfies the relation $\mathbf{m}_H^{(1)} = \alpha_m \mathbf{H}_{imp}$, in the quasi-static assumption that the magnetic field may still be considered uniform over the loop. Comparing (12) and (3) we find:

$$\alpha_m^{-1} = \frac{4\varepsilon_b}{N(k_b R)^2}\left[\alpha_p^{-1} - \sum_{j \neq j'}^{N-1} \mathbf{Q}_{jj'} \cdot \hat{\boldsymbol{\varphi}}(\mathbf{r}_j) \cdot \hat{\boldsymbol{\varphi}}(\mathbf{r}_{j'})\right], \tag{13}$$

where the term in the summation, being independent of $j'$, has expression given by (4). If we neglect the imaginary part of Eq. (13), its value becomes consistent with what we found in [26]. It is interesting to analyze more carefully in this context this imaginary part, using the Taylor expansion (5) and Eq. (8). We find, for any $N > 1$:

$$\mathrm{Im}\left[\alpha_m^{-1}\right] = \frac{4\varepsilon_b}{N(k_b R)^2}\left[-\alpha_{loss} - \sum_{j=1}^{N-1} \frac{k_b^5 R^2 \left[1 - 3\cos(2j\delta)\right]\sin^2(j\delta)}{30\pi\varepsilon_b} + o(k_b R)^4\right], \tag{14}$$

where the electric dipole contribution to radiation coming from $\alpha_p$, as given in (8), has been canceled by part of the summation [38]. The residual sum can be evaluated in closed forms as a function of the number of particles composing the nano-loop:

$$\begin{aligned}\mathrm{Im}\left[\alpha_m^{-1}\right] &= -\frac{4\varepsilon_b \alpha_{loss}}{N(k_b R)^2} - \frac{4k_b^3}{15\pi} + o(k_b R)^2 \quad \text{for } N = 2 \\ \mathrm{Im}\left[\alpha_m^{-1}\right] &= -\frac{4\varepsilon_b \alpha_{loss}}{N(k_b R)^2} - \frac{k_b^3}{6\pi} + o(k_b R)^2 \quad \text{for } N > 2\end{aligned} \tag{15}$$

These relations on the imaginary part of (14) are again related to energy conservation issues and describe how the power extracted by the nanoring from the impinging field is re-radiated in free space. In particular, since the power extracted by a magnetic dipole of amplitude $\mathbf{m} = \alpha_m \mathbf{H}_{imp}$ from an impressed magnetic field $\mathbf{H}_{imp}$ is equal to [27]:

$$P_{ext} = \frac{1}{2}\mathrm{Re}\left[i\omega\mu_0 \mathbf{m}^* \cdot \mathbf{H}_{imp}\right] = \frac{\omega\mu_0}{2}\mathrm{Im}\left[\alpha_m\right]\left|\mathbf{H}_{imp}\right|^2, \tag{16}$$

and the power radiated by such a magnetic dipole is equal to [39]:

$$P_{rad} = \frac{\sqrt{\mu_0/\varepsilon_b}k_b^4}{12\pi}|\mathbf{m}|^2 = \frac{\sqrt{\mu_0/\varepsilon_b}k_b^4}{12\pi}|\alpha_m|^2|\mathbf{H}_{imp}|^2, \qquad (17)$$

the following relation holds for a lossless magnetic dipole for which $P_{ext} = P_{rad}$:

$$\text{Im}\left[\alpha_m^{-1}\right] = -\frac{k_b^3}{6\pi}, \qquad (18)$$

which is analogous to the condition on the inverse electric polarizability of electric particles previously employed (8). When losses are present, the radiated power is less than the extracted one and a negative term $-\alpha_{mloss}$ is added to the right hand side of (18). Interestingly, in the limit of lossless particles in the loop (i.e., when $\alpha_{loss} = 0$), formula (15) confirms analytically the previous condition (18) for any $N > 2$, provided that we can neglect the higher-order terms $o(k_b R)^2$. This validates the dynamic theory derived here and ensures that a collection of three or more plasmonic nano-particles arranged around a loop indeed compose a pure magnetic molecule in the quasi-static limit. The additional contribution $o(k_b R)^2$ is associated with the spurious radiation from the higher-order multipoles, which is associated in the expression for the magnetic polarizability expression with radiation losses (since the extracted power does not coincide anymore with the power radiated by the magnetic dipole term). This contribution is negligible for sufficiently small loops and/or sufficiently high number $N$ of particles in the nanoring. For a pair of nanoparticles, i.e., when $N = 2$, however, the situation is drastically different: the expression in $\text{Im}\left[\alpha_m^{-1}\right]$ is modified in (15) by the inherent presence of the non-negligible electric quadrupole radiation, which is of the same order as the magnetic dipole moment, as previously noticed. In other words, for a nanopair the inevitable presence of electric quadrupole radiation adds an amount of radiation losses comparable with the magnetic dipole contribution, and the nanopair cannot be correctly interpreted as a lossless pure magnetic inclusion when embedded in a metamaterial lattice, even in the ideal case for which material absorption can be neglected. This contribution necessarily adds a negative term in (18), implying that a pair of plasmonic nanoparticles, even though supporting a resonant magnetic response, necessarily present a non-negligible radiation loss associated with their electric quadrupole radiation. This confirms the results of Fig. 2 and shows how *a pair of plasmonic nano-particles at their anti-symmetric resonance cannot be treated as a purely magnetic molecule*. A larger number of particles around the loop is necessary to compensate this effect and treat the molecule as a magnetic resonator, consistent with Eq. (15).

It is also worth noting how (15) implies that the ohmic absorption in each particle, related to $\alpha_{loss}$ has a weaker effect on the magnetic resonance when the number of employed particles is larger, since its effect is divided by $N$. From (15), we can indeed write the relation:

$$\alpha_{mloss} = \frac{4\varepsilon_b \alpha_{loss}}{N(k_b R)^2} + \frac{k_b^3}{10\pi} + o(k_b R)^2 \quad \text{for } N = 2$$

$$\alpha_{mloss} = \frac{4\varepsilon_b \alpha_{loss}}{N(k_b R)^2} + o(k_b R)^2 \quad \text{for } N > 2$$

(19)

The presence of ohmic losses in the particles is discussed more in details in the following section.

When larger particles are considered and the $o(k_b R)^2$ contribution cannot be neglected, its evaluation may be conducted by also considering a dynamic form for the excitation. In this case, the quasi-static assumption of a uniform magnetic field over the whole volume of the nanoring may indeed represent a too strong approximation and it should be substituted by a dynamic expression in the form $\mathbf{H}_{imp} = H_{imp} J_0(k_b r)\hat{\mathbf{z}}$, with $J_0(.)$ being the cylindrical Bessel function (this field distribution in fact satisfies Maxwell equations and ensures a zero-derivative symmetric magnetic field at the origin). In this way we find the dynamic expression for the impressed field on each particle, which substitutes (1):

$$\mathbf{E}_{imp} = i\sqrt{\frac{\mu_0}{\varepsilon_b}} H_{imp} J_1(k_b R)\hat{\boldsymbol{\varphi}}. \tag{20}$$

In this case the averaged magnetic field over the nano-loop is evaluated as:

$$\tilde{\mathbf{H}}_{imp} = \frac{\int_0^R \mathbf{H}_{imp} r\, dr}{R^2/2} = \frac{2 H_{imp} J_1(k_b R)}{k_b R}\hat{\mathbf{z}}, \tag{21}$$

and the magnetic polarizability, defined as $\mathbf{m}_H^{(1)} = \alpha_m \tilde{\mathbf{H}}_{imp}$, interestingly provides the same expression (13). It should be mentioned, however, that the same concept of polarizability and the corresponding homogenization procedure for a bulk metamaterial made of these nano-particles, as well as the quasi-static multipole expansion (9), lose part of their meaning when the size of the particle becomes too large and comparable with the wavelength of operation. The formulas derived here, therefore, hold until the higher-order multipole contributions indicated with $o(.)$ remain of secondary importance.

Having discussed how the imaginary part of (13) may provide an interpretation of the magnetic properties of the nanoring, we can analyze now how the real part of the inverse polarizability affects the nanoring design:

$$\operatorname{Re}\left[\alpha_m^{-1}\right] = \frac{4\varepsilon_b}{N(k_b R)^2} \operatorname{Re}\left[\alpha_p^{-1}\right] - \frac{k_b^{-2} R^{-5}}{16 N\pi} \sum_{j=1}^{N-1} \frac{3+\cos(2j\delta)}{\sin^3(j\delta)} + o(k_b R)^{-1}, \tag{22}$$

consistent with the quasi-static result in [26] (apart from a sign misprint in [26], Eq. (9)).

A magnetic resonance is achieved when $\text{Re}\left[\alpha_m^{-1}\right]=0$, which happens near the resonant frequency of each of the particle composing the loop (arising at $\text{Re}\left[\alpha_p^{-1}\right]=0$), but slightly shifted by the coupling term represented by the summation in (22). It is worth underlining that the magnetic resonance depends mainly on the resonant properties of each plasmonic particle composing the loop, rather than on the loop *geometry* or *size*, implying that a sub-wavelength magnetic resonance may be achieved, in principle independently of the total size of the nanoring. This is of particular importance for synthesizing sub-wavelength inclusions to be embedded in a metamaterial for homogenization purposes.

Once the magnetic polarizability of the nano-loop is evaluated, the effective permeability of a composite metamaterial made of an infinite 3-D lattice of such inclusions may be calculated using appropriate homogenizing formulas [28]. It should be underlined, however, that the possible presence of non-negligible spurious higher-order multipoles in the magnetic response of the nanoinclusion may sensibly affect this homogenization procedure, and it may be effectively embedded in the measured imaginary part of the effective $\mu$. For what discussed above, this is of special importance for nanopair inclusions ($N=2$). In the case of a regular cubic lattice, whose periodicity compensates the radiation losses due to the magnetic dipole radiation from each nanoring, the effective permeability calculated with a Clausius-Mosotti homogenization is given by [26]:

$$\mu_{eff}^{(p)} = \mu_0 \left(1+\left\{N_d^{-1}\left[\alpha_m^{-1}+i\left(k_0^3/6\pi\right)\right]-1/3\right\}^{-1}\right), \tag{23}$$

with $N_d$ being the number density of loops in the lattice. The previous formula implies that the imaginary part of $\mu_{eff}^{(p)}$ may be associated not only with the material absorption, but also with the radiation losses associated with higher-order multipoles.

It should also be mentioned that the presence of spurious multipolar radiation, particularly important for $N=2$, may give rise not only to radiation losses, but also to an unwanted spatial dispersion in the metamaterial response. If their contribution is not negligible, the assumption that the metamaterial response may be simply described by an effective permeability is not sufficient to properly describe its complex wave interaction. More specifically, in the metamaterials composed of nanopairs the presence of a non-negligible electric quadrupole contribution implies that the response to the symmetric part of the gradient of the local electric field should be properly embedded in modified constitutive relations [41], since the permeability factor by itself may neglect this response. In other words, it is indeed true that a nanopair may provide a resonant magnetic response at its anti-symmetric resonance, as shown in [14]-[22], but as the previous analysis shows, this effect is not separable from the comparable presence of electric quadrupole radiation and higher-order multipoles. Assuming that the metamaterial response may be described by a simple effective permeability may lead to inadequate results in terms of the prediction of its scattering losses and spatial dispersion. These side effects may also affect the quality factor of the magnetic resonance. These may be properly avoided by increasing the number of particles around the nano-loop.

Fig. 3 shows the frequency dispersion of the effective permeability $\mu_{eff}^{(p)}$ of a regular cubic lattice with number density $N_d = (160\,nm)^{-3}$ of nano-loops of radius $R = 60\,nm$ constituted of silver nanospheres with radius $a = 24\,nm$ in a free-space background. In the calculations, realistic ohmic losses and frequency dispersion of the silver have been considered using a classic Drude model [14]. From the figure, it is noticed that, when $N$ is increased, the permeability resonance is shifted by the larger mutual coupling among the particles, as predicted by (22), and that the magnetic resonance may be strongly enhanced by an increase in $N$. The case with $N = 2$ would create a relatively weak magnetic resonance around the antisymmetric resonant frequency of the pair, but the effective permeability would not yield negative values, being sensibly affected by ohmic and radiation losses. If the number of particles per loop is increased, the situation is drastically improved and the robustness to losses and bandwidth of operation is increased consistently.

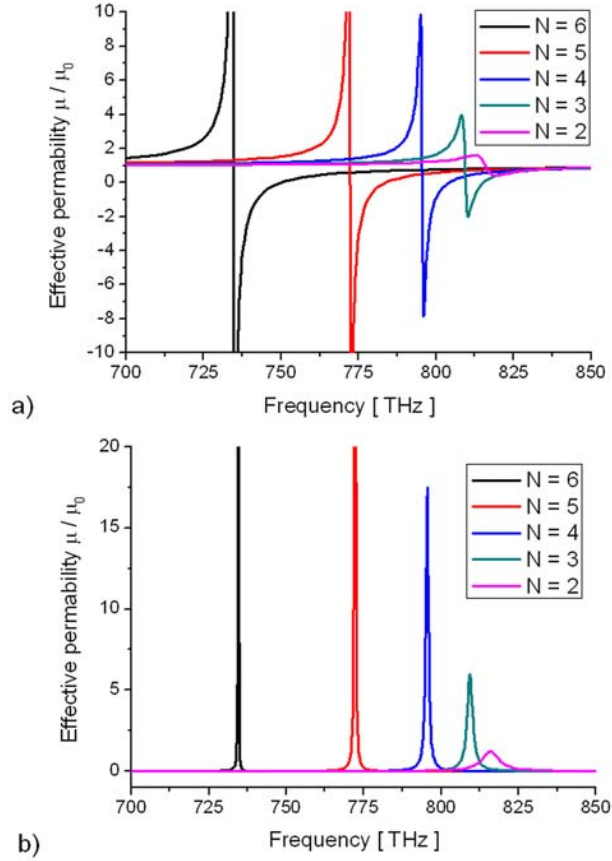

Figure 3 – (color online) (a) Real and (b) imaginary parts of the effective permeability for a cubic lattice with number density $N_d = (160\,nm)^{-3}$ made of nano-loops with radius $R = 60\,nm$ each of which is composed of $N$ silver particles with radius $a = 24\,nm$.

## 4. Losses and $Q$ factor

Here we analyze the behavior of this magnetic resonance as a function of the material losses. For this purpose, let us consider the case of nanospheres of radius $a$ and homogeneous permittivity $\varepsilon = \varepsilon_r + i\varepsilon_i$ composing the nanoring. In this case, the quasi-static expression for $\alpha_{loss}$ of each nanosphere is given by:

$$\alpha_{loss} = \frac{3}{4\pi} \frac{\varepsilon_i a^{-3}}{\left(\varepsilon_r - \varepsilon_b\right)^2 + \varepsilon_i^2}, \qquad (24)$$

which follows from the quasistatic expression of the electric polarizability of a homogeneous sphere [4]. This yields for the ohmic factor $\alpha_{mloss}$ the expression:

$$\alpha_{mloss} = \frac{k_b^3}{6\pi} \frac{18(k_b a)^{-3}}{N(k_b R)^2} \frac{\varepsilon_b \varepsilon_i}{\left(\varepsilon_r - \varepsilon_b\right)^2 + \varepsilon_i^2}. \qquad (25)$$

As expected, this factor is proportional in the low-loss limit to the quantity $\varepsilon_i$. Moreover, a scaling up of the nanoring and nanoparticle size, or equivalently a reduction of the frequency of operation, decreases the sensitivity to the material losses. This represents the main lower limit on the possibility of squeezing the dimensions of such magnetic nanorings, since the sensitivity to the material losses increases with a size reduction. On the other hand, a more negative real part of permittivity for each particle at the frequency of operation allows reducing this sensitivity, due to the corresponding reduction in the wave penetration (skin depth) in each of the nanoparticles. This reduction factor is higher than the loss tangent in the material (as it has been speculated for instance in [20]) and it is represented with good approximation by the ratio $\varepsilon_i / \left(\varepsilon_r - \varepsilon_b\right)^2$. For a fixed geometry of the nanoparticles, a larger number of particles composing the loop shifts down the resonant frequency, as shown by (22), and generally at lower frequencies plasmonic materials have a lower negative permittivity $\varepsilon_r$, helping this phenomenon. Varying the geometry of each nano-particle composing the loop in order to move their individual resonance to a more negative permittivity value may also help in reducing the effect of losses. Also, increasing the background permittivity $\varepsilon_b$ may proportionally increase the absolute value $-\varepsilon_r$ at which the loop has its resonance, reducing its sensitivity to losses. Finally, a larger number of nanoparticles composing each loop reduces directly the radiation losses, as already described, since $N$ is in the denominator of (25).

The $Q$ factor of this magnetic resonance may be also evaluated in closed form, providing an insight into its expected bandwidth of operation. For the case of a ring constituted by plasmonic spheres of radius $a$ and Drude permittivity $\varepsilon = \left(1 - 3\omega_0^2 / \omega^2\right)\varepsilon_b$ (here the effect of losses is neglected), the individual resonance of each particle arises at the angular frequency $\omega_0$. The overall magnetic resonance is obtained, using (22), for:

$$\omega_{m0} = \omega_0 \sqrt{1 - \frac{a^3}{16R^3} \sum_{j=1}^{N-1} \frac{3 + \cos(2j\delta)}{\sin^3(j\delta)}}, \tag{26}$$

shifted down in frequency with respect to the self-resonance of each nano-particle due to the coupling among the nanoparticles. It is interesting to note, as an aside, how the shifting factor under the square root satisfies the inequality $0 < \frac{a^3}{16R^3} \sum_{j=1}^{N-1} \frac{3 + \cos(2j\delta)}{\sin^3(j\delta)} < \frac{\zeta(3)}{2} \simeq 0.6$, since the geometrical constraint $a \leq R\sin(\pi/N)$ holds [42].

The $Q$ factor of this resonance is readily evaluated from the previous formulas and it is equal to the following expression in this quasi-static limit:

$$Q = \frac{6\left(1 - \frac{a^3}{16R^3} \sum_{j=1}^{N-1} \frac{3 + \cos(2j\delta)}{\sin^3(j\delta)}\right)}{N\tilde{k}_b^5 a^3 R^2}, \tag{27}$$

where $\tilde{k}_b$ is the background wave number calculated at $\omega = \omega_{m0}$.

It is evident that the $Q$ factor of the magnetic resonance of the nano-loop of Fig. 1 increases with a reduction of its radius, with a reduction of the size of the nano-particles composing it, and with an increase of the number of particles in the loop, consistently with the previous discussion. The fractional bandwidth of operation is simply given by $1/Q$.

The electric response of the nano-loop of Fig. 1 may be evaluated numerically as derived in [26], yielding a resonance generally very close to the individual resonance of each of the nanoparticles composing the loop. A judicious design of the geometry of Fig. 1 may give rise to overlapping electric and magnetic resonances, effectively providing a way to design a left-handed metamaterial at optical frequencies. We reiterate the importance of employing multiple nanoparticles in the design of the nanoring in order to increase the associated magnetic effects and reduction of losses and spatial dispersion.

## 5. Conclusions

We have investigated in detail the physics underlying the magnetic resonance of plasmonic nanorings at infrared and optical frequencies, consistent with the geometry presented in [26]. In particular, we have proven analytically and numerically the superiority of the nanoring geometry over pairs of nanoparticles, due to a purer magnetic response, necessary to produce left-handed materials with lower absorption. Analytical closed-form expressions for the sensitivity to losses and the $Q$ factor and bandwidth of the associated magnetic resonance have also been derived and discussed.

**Acknowledgments**

This work is supported in part by the U.S. Air Force Office of Scientific Research (AFOSR) grant number FA9550-05-1-0442. The authors thank Prof. Mario G. Silveirinha for some useful discussions.

[34] From (9) and geometric considerations it also follows that the electric multipoles of order $N+1$ are identically zero and, for even $N$, also all the odd electric multipoles vanish.